\documentclass[twocolumn,amsmath,amssymb,pre]{revtex4}
\usepackage{textcomp}
\usepackage{amsmath}
\usepackage{amssymb}
\usepackage{graphicx}
\usepackage{epstopdf}
\usepackage{bbm}
\makeatletter

\usepackage{textcomp}
\usepackage{epstopdf}
\usepackage{bbm}
\usepackage[toc,page,title,titletoc,header]{appendix}
\usepackage{dcolumn}\usepackage{bm}
\newcommand{\be}{\begin{equation}}
\newcommand{\ee}{\end{equation}}
\newcommand{\bey}{\begin{eqnarray}}
\newcommand{\eey}{\end{eqnarray}}
\newcommand{\bw}{\begin{widetext}}
\newcommand{\ew}{\end{widetext}}
\newcommand{\ww}{\widetilde}

\newcommand{\ra}{\rangle}
\newcommand{\la}{\langle}

\newcommand{\E}{{\cal E}}

\newcommand{\ba}{\begin{array}}
\newcommand{\ea}{\end{array}}
\newcommand{\bi}{\begin{itemize}}
\newcommand{\ei}{\end{itemize}}
\newcommand{\bem}{\begin{enumerate}}
\newcommand{\eem}{\end{enumerate}}

\makeatother

\begin{document}

\title{A decoherence interpretation of quantum work for adiabatic processes }

\author{Wen-ge Wang \footnote{ Email address: wgwang@ustc.edu.cn}}

\affiliation{Department of Modern Physics, University of Science and Technology
of China, Hefei, 230026, China}

\date{\today}
\begin{abstract}
In this paper, decoherence is studied for quantum systems undergoing adiabatic processes,
 which are coupled to huge quantum environments. It is shown that decoherence can happen 
with respect to a preferred basis given by transient self-Hamiltonian.
This result can be used to justify a most-often-used definition of quantum work for a system
undergoing an adiabatic process, without resorting to measurement.
\end{abstract}

%\pacs{05.30.-d, 07.20.Dt, 05.45.Mt, 06.20.-f  }

%05.30.-d Quantum statistical mechanics
%05.45.Mt Quantum chaos; semiclassical methods
%05.90.+m Other topics in statistical physics, thermodynamics, and nonlinear dynamical systems (restricted to new topics in section 05)
%06.20.-f Metrology
%07.20.Dt Thermometers
%34.10.+x General theories and models of atomic and molecular collisions and interactions

\maketitle

\section{Introduction}

 Quantum work is a notion, which has attracted lots of attentions and investigations in recent years
 (see, e.g., Ref.\cite{Hanggi16} and references therein).
 A most-often-used definition is via two-point energy measurements,
 in which projective energy measurements are performed 
 at the initial and final times for a driving process, and the quantum work is computed by the difference
 of the measured energies at the two times  \cite{Kurchan07,Tasaki09,TLH07}.
 A merit of this definition is that the statistics of work complies with the fluctuation theorems of 
 Jarzynski \cite{Jarzy97} and Crooks \cite{Crooks99}.
 Recently, this definition is further justified from the angle of 
 the quantum-classical correspondence principle
 in integrable \cite{Quan1}, chaotic \cite{Quan2}, and many-body systems \cite{Quan3}.

 However, in the conceptual aspect, further justification is still needed
 for the use of projective measurements, which involves collapse of state vectors
 and may lead to interpretational difficulty.
 In fact, in a intuitive understanding of thermodynamic work, 
 the work should be computable from properties of a given process,
 independent of whether the work is really measured or not.  
 A crucial point is that, unlike in classical mechanics, 
 a measurement on a quantum system usually modifies the state of the system and,
 as a result, it may change the subsequent evolution of the system.

 To get a deeper understanding of quantum work, one may consider the evolution of a total system,
 including the studied system and its environment.
 Indeed, a driven system must be driven by some part of its environment and a measurement on it must
 be performed by a device in its environment. 
 A key point lies in a phenomenon  well-known
 for open quantum systems, namely, decoherence \cite{Zurek03,Schloss05}.
 In fact, when decoherence has happened such that the reduced density matrix (RDM) of the studied system
 becomes diagonal in the eigenbasis of itself Hamiltonian \cite{Zurek-ps,pra08-ps},
 the RDM has a mixture form and supplies the same information about
 the energy of the system as that obtained from projective energy measurements. 
% for which previous coherence effects with respect to this basis 
% (if ever existed) can not resume due to complex motion of the environment 
% during any finite time period of practical interest.

 In this paper, we show that the above-discussed decoherence can indeed happen
 for a driven system undergoing an adiabatic process, with the total system undergoing  
 a Schr\"{o}dinger evolution.
 For such an adiabatic process, since there is no heat transfer, the energy change can be
 identified with the work done. 
 Then, the quantum work can be computed, without resorting to measurements,
 and the result coincides with that given by the two-point energy-measurement definition discussed above.
 The paper is organized as follows. In Sec.\ref{sect-setup}, we discuss the setup of the problem,
 in particular, we write the total Hamiltonian in a form appropriate for the 
 study of adiabatic processes. 
 Decoherence effects are discussed in Sec.\ref{sect-decoh} and conclusions are given in 
 Sec.\ref{sect-conclu}.

\section{The form of the total Hamiltonian for adiabatic process}\label{sect-setup}

 We use $S$ to denote the studied system, which is coupled to a huge environment denoted by $\E$.
 The total system is governed by a Hamiltonian, 
\begin{gather}\label{}
 H = H_S + H_I + H_\E, \\  H_I = H_I^S \otimes H_I^\E ,
\end{gather}
 where for the simplicity in discussion we have assumed that the interaction Hamiltonian $H_I$ has a 
 direct-product form. 
 Schr\"{o}dinger evolution of the total system is written as $|\Psi(t)\ra = e^{-iHt}|\Psi(0)\ra$.
 The corresponding density operator is written as $\rho = |\Psi\ra \la \Psi|$. 
  The system $S$ is described by its RDM, denoted by $\rho^S$,  $\rho^S \equiv {\rm Tr}_\E \rho$.

% One should note that, for a give system and a given environment, the total Hamiltonian should be
% fixed, which can in principle be obtained from the four basic interactions known for the nature. 

 We assume that the environment can be divided into two parts, denoted by $\E_1$ and $\E_2$.
 The part $\E_1$ is small and undergoes a regular motion,
 for which the expectation value of certain observable (such as the position of a piston) can be
 regarded as a controlling parameter for the system $S$.
 While, the part $\E_2$ is a huge, quantum chaotic system, as the main body of the environment. 
 Accordingly, $H_I^\E$ and $H_\E$ are written as
\begin{gather}\label{}
 H_I^\E  = H_I^{\E_1} \otimes H_I^{\E_2}, \\
 H_\E = H_{\E_1} \otimes I_{\E_2} +  I_{\E_1} \otimes H_{\E_2},
\end{gather}
 where $I_{\E_1}$ and $I_{\E_2}$ indicate identity operators in the subspaces for 
 the two parts of the environment. 
 We assume that the chaotic motion of $\E_2$ has negligible influence in the motion of $\E_1$,
 such that the interaction between the two parts of the environment can be neglected in our study.

 As well known, under the same total Hamiltonian, there is some freedom in choosing the form of the 
 self-Hamiltonian of the system $S$ and the interaction Hamiltonian. 
 In the study of decoherence, the most appropriate division of these two parts of the total Hamiltonian
 can be made, according to the so-called preferred (pointer) basis of the system (if such a basis exists), 
 in which the RDM keeps a diagonal form \cite{Zurek-ps,pra08-ps}.
 Below, we discuss this division and use it to determine a controlling parameter
 for an adiabatic process of the system $S$.

 A system, which initially lies in an equilibrium state and undergoes an adiabatic process,
 should remain in a quasi-equilibrium state during the time evolution.
 This quasi-equilibrium state is approximately a steady state. 
 Preferred basis for steady states has been studied in Ref.\cite{pre14-ps}.
 There, it is shown that, for a system coupled to a huge, chaotic quantum environment, 
 if ${\rm Tr}_\E (H_I) =0$, then, a steady state $\rho^S$ should be commutable with the self-Hamiltonian $H_S$, 
 with an error scaling as $N_\E^{-1/2}$, where $N_\E$ is the dimension of the 
 Hilbert space of the environment. 
 In the case that $\rho^S$ is not close to the identity operator,  this result implies that 
 the eigenbasis of the self-Hamiltonian should supply a preferred basis.

 In the derivation of the above result, the quantum motion of the environment is assumed
 to explore the whole energy region.
 In the situation considered here, the quantum motion of the
 chaotic part of the environment is usually
 restricted within a narrow energy region and, hence,
 within a subspace of its Hilbert space.
% Due to the hugeness of the environment,
 It is reasonable to assume that the dimension of this subspace, denoted by
 $N_{\ww \E_2}$, is much larger than the 
 dimension of the Hilbert space of the system $S$.
 We use ${\rm Tr}_{\ww\E_2} (H_I^{\E_2}) $ to denote the trace of $H_I^{\E_2}$
 in this subspace.
 Then, a generalization of the above-discussed result of Ref.\cite{pre14-ps} states that,
 when ${\rm Tr}_{\ww\E_2} (H_I^{\E_2}) =0$,
 the RDM $\rho^S$ should be commutable with $H_S$
 with an error scaling as $N_{\ww\E_2}^{-1/2}$.

 In a generic situation, one may have ${\rm Tr}_{\ww\E_2} (H_I^{\E_2}) \ne 0$.
 This generic case can be treated by a method discussed in Ref.\cite{pre14-ps}, 
 in which a renormalized self-Hamiltonian, denoted by $H_S^r$, and a renormalized interaction Hamiltonian, 
 denoted by $H_I^r$, are introduced, with the total Hamiltonian unchanged.
 Specifically, 
\begin{gather}
 H_S^r = H_S + H_I^S \la H_I^{\E} \ra , 
 \\ H_I^r =  H_I^S \otimes (H_I^{\E} - \la H_I^{\E} \ra ),
\end{gather}
 where $\la H_I^{\E} \ra $ indicates certain average of $H_I^{\E} $, 
 which is equal to ${\rm Tr}_\E (H_I^\E) $ in the case discussed in Ref.\cite{pre14-ps}.

 To study an adiabatic process, we assume that $H_I^{\E_1}$ for the regular part 
 of the environment can be replaced approximately by its expectation value;
 for example, in the case of a piston, it may be given by the expectation value of its position.
 We use $\lambda(t)$ to indicate this expectation value.
 Then, $\la H_I^{\E} \ra$ can be written as $\la H_I^{\E} \ra = \lambda(t) \la  H_I^{\ww\E_2} \ra$,
 where $\la H_I^{\ww\E_2} \ra \equiv  {\rm Tr}_{\ww\E_2} (H_I^{\E_2})$. 
 As a result, the total Hamiltonian is written as 
\begin{gather}
   H(t) = H_S^r(t) + \lambda(t) H_I^{r2}(t)  + H_\E ,
\end{gather}
 where
 \begin{gather}
 H_S^r(t) = H_S +\lambda(t) \la H_I^{\ww\E_2} \ra H_I^S  , 
 \\ H_I^{r2} =  H_I^S \left[ H_I^{\E_2} - \la H_I^{\ww\E_2} \ra \right].
 \end{gather}

 In the specific case that the interaction term $\lambda(t)H_I^{r2}(t)$ is sufficiently weak 
 and has negligible influence in the evolution of the system $S$,
 one may neglect this term and consider Schr\"{o}dinger evolution of $S$ under a Hamiltonian $H_S^r$.
 If, furthermore, the value of $\lambda$ changes very slowly with the time $t$, 
 the system $S$ will undergo an ordinary adiabatic process.
 However, generically, the term $\lambda(t)H_I^{r2}(t)$
 can not be neglected, particularly when discussing decoherence effects.
 In this case, in the study of an adiabatic process of the system $S$,
 we still assume that the term $\lambda(t) H_I^{r2}$ be quite weak.

\section{Decoherence during an adiabatic process}\label{sect-decoh}

 In this section, we show that decoherence may happen with respect to the eigenbasis of $H_S^r$. 
 Let us consider the time evolution of the total system within two fixed times $t_0$ and $t_1$,
 between which $\lambda(t)$ changes slowly. 
 We divide the total Hamiltonian into two parts, an unperturbed part $H_0$ and a 
 perturbation $\epsilon H_1$,
\begin{gather}\label{}
 H = H_0 + \epsilon H_1,
\end{gather}
 where 
\begin{gather}\label{}
 H_0 = H_S^r(t_0)  + H_\E, \quad
 \\ \epsilon H_1 =  \Delta H_S + \lambda(t) H_I^{r2}.
\end{gather}
 Here, 
\begin{gather}\label{}
 \Delta H_S = H_S^r(t) -H_S^r(t_0) = (\lambda(t)-\lambda(t_0))  \la H_I^{\E_2} \ra H_I^S.
\end{gather}
 As discussed above, $\lambda(t) H_I^{r2}$ is a quite weak perturbation, meanwhile,
 due to the slow variation of $\lambda$, $\Delta H_S$ is also small.
 Hence, $\epsilon H_1$ is a quite weak perturbation
 and it proves convenient to set $\epsilon$ a small parameter.

 When the variation rate, $d\lambda /dt $, is sufficient small,  
 the term $\Delta H_S$ can be neglected for a given period $[t_0,t_1]$ and the perturbation can be written 
 in a time-independent form, 
\begin{gather}\label{}
 \epsilon H_1 = \lambda(t_0) H_I^{r2},
\end{gather}
 where we have neglected possible time-dependence of $\la H_I^{\ww\E_2} \ra$.
 Let us use $|\alpha\ra$ to denote eigenstates of $H_S^r(t_0)$, 
\begin{gather}\label{}
 H_S^r(t_0)|\alpha\ra = E_\alpha |\alpha\ra.
\end{gather}
% and, for the simplicity of discussion, assume that the spectrum $E_\alpha$ is nondegenerate. 
 Elements of the RDM in this basis are written as $\rho^S_{\alpha\beta} \equiv \la\alpha|\rho^S|\beta\ra$.

 Suppose that the perturbation $\epsilon H_1$ is sufficiently weak such that,
 in the perturbation expansion for the transition among the states $|\alpha\ra$,
 contributions from the second and higher-order terms can be neglected.
 In this situation, results  given in Ref.\cite{pra08-ps} are directly applicable.
 To present these results, it proves convenient to introduce an operator $V$ 
 in the Hilbert space for the environment, related to two states $|\alpha\ra$ and 
 $\beta$ of the system $S$, namely,
\begin{gather}\label{V-define}
 V \equiv \la\beta|H_1|\beta \ra - \la \alpha|H_1|\alpha\ra.
\end{gather}
 A perturbative border $\epsilon_p$ is of relevance,
 which can be estimated by the following relation,
 \be 2\pi \epsilon_p \overline{ V_{nd}^2} \sim \sigma_v \Delta , \label{per-bor} \ee
 where  $\overline{ V_{nd}^2}$ indicates the average of $|\la n|V|n'\ra |^2$
  with $n \ne n'$, $\Delta $ is the mean level spacing of $H^{eff}_{\E \alpha }$,  defined by
\begin{gather}\label{Heff}
 H^{eff}_{\E \alpha } = H_\E + \epsilon \la \alpha|H_1|\alpha\ra,
\end{gather}
 and $\sigma_{v}^2$ is the variance of $\la n|V|n\ra $.
 Here, $|n\ra $ denotes eigenstates of $H^{eff}_{\E \alpha }$ \cite{foot-V}.

 For a weak perturbation such that $\epsilon < \epsilon_p$, as discussed in Ref.\cite{pra08-ps}, 
\begin{gather}
 |\rho^s_{\alpha \beta }(t)| \sim \exp\left(-\epsilon^2 \sigma^2_{v}t^2/2 \hbar^2\right), \label{Gaussian} 
\end{gather}
 which gives a decoherence rate $R_{d}$, 
 \be \label{tau-f1} R_d \simeq \epsilon \sigma_{v}/(  \sqrt{2} \hbar)   \propto \epsilon.
\ee
 Meanwhile, the transition between the states $|\alpha\ra$ can be estimated by Fermi's golden rule.
 Let $|\mu_{\E}\ra $ be eigenstates of $H_{\E}$
 and $ \la H^2_{1,nd} \ra $ be the mean square of the non-diagonal matrix elements
 $\la \alpha '|\la \mu_{\E}'| H_1 |\mu_{\E}\ra |\alpha \ra $ ($\alpha\ne\alpha'$).
 Then, Fermi's golden rule predicts that the transition rate $R_E$ is given by
 \be  R_E = 2\pi \epsilon^2 \rho_E \la H^2_{1,nd} \ra / \hbar \propto \epsilon^2, \label{te-fgr} 
\ee
 where $\rho_E$ indicates the density of states of the total system.

 For sufficiently small $\epsilon$, Eqs.(\ref{tau-f1})-(\ref{te-fgr}) show that $R_d \gg R_E$,
 that is, the decoherence rate is much larger than the transition rate.
 As a consequence, whenever the RDM may have nonnegligible off-diagonal elements
 in the eigenbasis of the Hamiltonian $H_S^r(t)$, 
 these off-diagonal elements will decay to very small values, 
 before notable transition among the states $|\alpha\ra$ shows up. 
 Therefore, for an adiabatic process, if initially the RDM of the system $S$ 
 has a diagonal form in the eigenbasis of $H_S^r(0)$,
 it will keep an approximately-diagonal form in the eigenbasis of $H_S^r(t)$;
 if initially the RDM has nonnegligible off-diagonal elements in the eigenbasis of $H_S^r(0)$, these elements
 will decay to quite small values and then the RMD will keep an approximately-diagonal form
  in the eigenbasis of $H_S^r(t)$. 

 The above results imply that, if initially the RDM of the system $S$ is described by a Gibbs state, then,
 for later times of an adiabatic process, the RDM should be described by a mixture, 
 in which the system lies in a state $|\alpha(t)\ra$ with a probability $\rho^S_{\alpha \alpha} $,
 where $|\alpha(t)\ra$ indicate eigenstates of $H_S^r(t)$.
 It is clear that the energy change computed from this mixture
 at difference times can be interpreted as quantum work,
 as assumed in the definition of quantum work discussed in the beginning of this paper.

\section{Conclusion}\label{sect-conclu}

 To summarize, in this paper it is shown that for a quantum system undergoing an adiabatic process,
 which is coupled to a huge quantum environment, 
 decoherence can happen with respect to a preferred basis given by the eigenbasis of its
 transient self-Hamiltonian.
 This result can be used to justify a most-often-used definition of quantum work done during an 
 adiabatic process, without resorting to measurement.

 It is not difficult to see that the method discussed in this paper is also useful for a more generic process,
 which is composed of some adiabatic parts and some nonadiabatic parts.
 Decoherence may happen within the adiabatic parts of the process (if they are sufficiently long) 
 and quantum work can be computed in a similar way for these parts.
 But, for times within the nonadiabatic parts, further investigation is needed
 in order to know whether the method discussed in this paper can be useful.

% The authors are grateful to for valuable discussions and suggestions.
 This work was partially supported by the Natural Science Foundation of China under Grant
 Nos.~11275179 and 11535011,
 and the National Key Basic Research Program of China under Grant
 No.~2013CB921800.


\begin{thebibliography}{1}
\bibitem{foot-V} We remark that, if $V$ involves only a small part of the quantum chaotic environment, 
 which is the case in a generic situation, the quantity $\overline{ V_{nd}^2}$
 scales as $1/N_\E$, meanwhile, $\Delta $ also scales as $1/N_\E$.
\end{thebibliography}
\end{document}